\documentclass[journal]{IEEEtran}

\usepackage{amsmath}

\ifCLASSINFOpdf
  \usepackage[pdftex]{graphicx}
\else
  \usepackage[dvips]{graphicx}
\fi

\usepackage{array}
\usepackage[printonlyused]{acronym}
\usepackage{cite}
\usepackage[tight]{subfigure}
\usepackage{multirow}

\usepackage{color, soul}
\usepackage{upgreek}

\newcommand{\noindento}{\vspace{0.3cm} \noindent}

\title{Medium Access Control in Wireless Network-on-Chip: A Context Analysis}

\author{Sergi~Abadal,
			Albert~Mestres,
			Josep~Torrellas,
        Eduard~Alarc\'on,
        Albert~Cabellos-Aparicio
\thanks{All authors except Josep Torrellas are with the NaNoNetworking Center in Catalonia (N3Cat),
    Universitat Polit\`ecnica de Catalunya, Barcelona, Spain.
    Josep Torrellas is with the i-acoma group, 
    University of Illinois, Urbana-Champaign, IL, USA.
    E-mail: \{abadal,amestres,acabello\}@ac.upc.edu, torrella@illinois.edu, eduard.alarcon@upc.edu}}

\acrodef{RC}{Resistive-Capacitive}
\acrodef{CMOS}{Complementary Metal--Oxide--Semiconductor}
\acrodef{mm-Wave}{millimeter-Wave}
\acrodef{WNSN}{Wireless NanoSensor Network}
\acrodef{WSN}{Wireless Sensor Network}
\acrodef{MAC}{Medium Access Control}
\acrodef{QoS}{Quality of Service}
\acrodef{TS-OOK}{Time Spread On-Off Keying}
\acrodef{CSMA}{Carrier Sense Multiple Access}
\acrodef{GWNoC}{Graphene Wireless Network-on-Chip}
\acrodef{WNoC}{Wireless Network-on-Chip}
\acrodef{NoC}{Network-on-Chip}
\acrodef{CMP}{Chip Multiprocessor}
\acrodef{TDMA}{Time Division Multiple Access}
\acrodef{FDMA}{Frequency Division Multiple Access}
\acrodef{CDMA}{Code Division Multiple Access}
\acrodef{ACK}{Acknowledgment message}
\acrodef{RF}{Radio-Frequency}
\acrodef{IR}{Impulse Radio}
\acrodef{OOK}{On-Off-Keying}
\acrodef{BER}{Bit Error Rate}
\acrodef{DVFS}{Dynamic Voltage and Frequency Scaling}
\acrodef{ARQ}{Automatic Repeat reQuest}
\acrodef{FEC}{Forward Error Correction}
\acrodef{M2M}{Machine-to-Machine}
\acrodef{mmWave}{millimeter-wave}

\begin{document}

\maketitle

\begin{abstract} 
Wireless on-chip communication is a promising candidate to address the performance and efficiency issues that arise when scaling current \ac{NoC} techniques to manycore processors. A \ac{WNoC} can serve global and broadcast traffic with ultra-low latency even in thousand-core chips, thus acting as a natural complement of conventional and throughput-oriented wireline NoCs. However, the development of \ac{MAC} strategies needed to efficiently share the wireless medium among the increasing number of cores remains as a considerable challenge given the singularities of the environment and the novelty of the research area. In this position paper, we present a context analysis describing the physical constraints, performance objectives, and traffic characteristics of the on-chip communication paradigm. We summarize the main differences with respect to traditional wireless scenarios, to then discuss their implications on the design of \ac{MAC} protocols for manycore \acp{WNoC}, with the ultimate goal of kickstarting this arguably unexplored research area. 
\end{abstract}






\acresetall

\section*{\Large \textbf{Introduction}} 

The relentless march of technology scaling has forced a widespread transition in processor design from \emph{single core} to \emph{multicore} due to power and complexity reasons. After this paradigm shift, successive generations of processors have changed the way to achieve higher performance from increasing the operation frequency to integrating more cores within the same chip. Research and industry are currently pushing this trend towards what has been called the {\em manycore era}, with predictions pointing towards chips with a thousand cores \cite{Borkar2007}. 


The path leading to the manycore era has gradually turned intra-chip communications, which occurs between the cores and the memory, into a key determinant of the computational performance and energy efficiency of multicore processors. Consequently, the on-chip interconnect has been the focus of a very large amount of research in recent years. Buses have given way to the more scalable, more efficient, and more resilient \ac{NoC} paradigm consisting of a fabric of on-chip routers and \ac{RC} wires. This conventional definition of \ac{NoC} works reasonably well for a moderate number of cores; however, intrinsic delay and power consumption limitations suggest that alternative approaches may be required in the manycore scenario.

In response to these shortcomings, several works have proposed to use new interconnect technologies in \acp{NoC} \cite{Kim2012Survey}. Wireless on-chip communication, enabled by recent advances in integrated \ac{RF} design, is among the considered alternatives given its multiple advantages:

\begin{itemize}
\item Low latency for communications between distant cores by virtue of speed-of-light propagation,
\item Natural broadcast capabilities via omnidirectional radiation,
\item System-level flexibility and non-intrusiveness given by the lack of additional wires between cores,
\item Compatibility with \ac{CMOS} and reuse of knowledge gained in other wireless scenarios.
\end{itemize}

\begin{figure}
  \centering
  \includegraphics[width=\columnwidth]{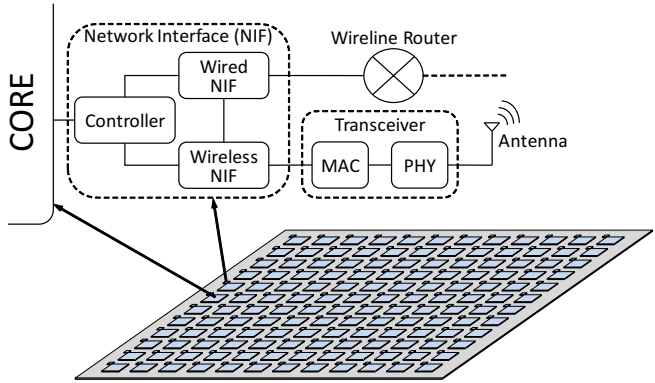}
  \vspace{-0.15cm}
	\caption{Schematic diagram of a manycore processor integrating a wireline NoC and a WNoC.}
  \label{fig:gwnoc}
  \vspace{-0.5cm}
\end{figure}

Integrated transceivers and antennas in the \ac{mmWave} band offer potential for transmission speeds of up to several tens of Gb/s, enough to transfer a 64B cache line {\em anywhere} on the chip in 5--15 cycles. This transparency of data transmission with respect to the location of data is extremely appealing in this application context, yet very difficult to achieve with current interconnects. Different works propose to exploit such competitive advantage to greatly reduce the latency and energy of transmissions between far-apart cores \cite{Deb2013, DiTomaso2015}, or to create a fast and efficient broadcast plane that opens a large design space from the architecture perspective \cite{AbadalMICRO}. In either case, \acp{WNoC} are not expected to completely replace wired interconnects, but rather to complement them. Such hybrid approach, illustrated in Figure \ref{fig:gwnoc}, consists in the integration of wireless links in a set of routers or network interfaces. Then, the routing protocol is modified so that long-range and broadcast flows are served by the wireless network and the rest of communications are served by the wired interconnect.


Harnessing the performance gains mentioned above requires addressing outstanding challenges at different levels of design. Matolak {\em et al} describe several of these challenges and review research initiatives relevant to transceiver design, channel modeling, and network architecture in \ac{WNoC} \cite{Matolak2012}. This article focuses on the issue of \ac{MAC} instead, a functionality that becomes essential as future \acp{WNoC} are expected to integrate tens of wireless interfaces that will share a small set of frequency channels.  

Related works on \ac{MAC} for \ac{WNoC} can be divided into two groups. On the one end, most existing \ac{WNoC} proposals strongly rely on the creation and distribution of fixed orthogonal channels \cite{Kim2012Survey}. These techniques are free of wasteful collisions and are capable of delivering high throughput, but do not work well under variable workloads since bandwidth is statically allocated. As explained later, channelization techniques also have important scalability limitations as increasing the number of time, frequency, or code-multiplexed channels imply a notable increase of the hardware complexity. The use of coordinated access protocols such as token passing, either alone \cite{Deb2013} or in combination with channelization techniques \cite{DiTomaso2015}, alleviates these flexibility and scalability concerns. Still, the overhead associated to the token circulation may hinder the use of such technique in manycore environments \cite{AbadalTPDS}. 

On the other end, random access techniques provide flexible operation and low latency as nodes can attempt to gain access at any time instant. However, this comes at the risk of collisions, overlapping transmissions that are discarded and that negatively impact on the energy efficiency and throughput of the network. Few works have investigated the use of random access in the \ac{WNoC} scenario. For instance, Abadal {\em et al} provide a comparison between token passing and \ac{CSMA} in a wide set of configurations \cite{AbadalTPDS}, confirming the advantages and disadvantages of the random access approach. Further, Mansoor {\em et al} propose a hybrid protocol that dynamically switches between \ac{CSMA} and token passing with the aim to achieve low latency and high throughput \cite{Mansoor2015}. 


This paper builds on the observation that the multiprocessor scenario represents a new environment for \ac{MAC} protocol research. As a result, the pioneering proposals mentioned above may be suboptimal as multiprocessors scale and the pressure upon the \ac{MAC} layer grows. In the quest for high performance and efficiency, \ac{MAC} solutions in manycore \acp{WNoC} will need to be fully aware of the physical constraints, performance objectives, and traffic characteristics unique to the on-chip communication paradigm. In fact, achieving near-optimal performance may require a careful review of the knowledge gained in other wireless scenarios and a complete rethinking of existing techniques.


In this article we aim to bridge this conceptual gap by providing, as the main contribution, a rigorous context analysis that details the main particularities of the on-chip scenario. Then, we discuss the possible impact that these uniquenesses may have on the design of MAC protocols for manycore \acp{WNoC}, as well as the challenges that can be found along the way, with the ultimate goal of kickstarting this research area.

\section*{\Large \textbf{Context Analysis}} 
Table \ref{tab:reqs} provides a rough quantification of the main requirements of the wireless manycore scenario. The number of nodes per transmission range reaches levels commensurate to those of massive \ac{WSN} or \ac{M2M} networks \cite{Niu2015}. The throughput demands also lead to strong resemblances with \ac{M2M} networks, including the use of \ac{mmWave} technologies. Finally, the on-chip networking scenario shares with mission-critical \acp{WSN} the need for latency and reliability guarantees, although with much more restrictive deadlines and power budget \cite{Suriyachai2012}.

\begin{table}[!t] 
\caption{Wireless Manycore Scenario Requirements}
\label{tab:reqs}
\footnotesize
\centering
\begin{tabular}{ll} 
\hline
{\bf Metric} & {\bf Value} \\
\hline
Transmission Range & 0.1--10 cm \\
Node Density & 10--1000 nodes/cm\textsuperscript{2} \\
Network Throughput & 10--100 Gb/s \\
Latency & 1--100 ns \\
Bit Error Rate (BER) & 10\textsuperscript{-15} \\
Energy & 1--10 pJ/bit \\
\hline
\end{tabular}
\vspace{-0.3cm}
\end{table}

Such a distinctive combination of requirements would be unsolvable in the aforementioned scenarios due to a long list of design issues: unknown topology, intermittent nodes, blockage, multipath, energy preservation constraints, or deafness problems, to name a few. However, we will see that the uniquenesses of the on-chip scenario virtually eliminate most of these problems, thus pointing towards simple and streamlined solutions through informed design decisions. 

\begin{figure}
  \centering
  \includegraphics[width=0.95\columnwidth]{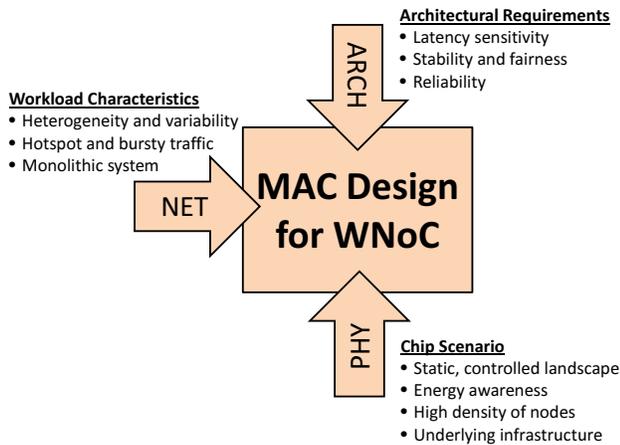}
  \caption{Main facets that define the MAC design process in a \ac{WNoC}.}
  \label{fig:MACscenario}
   \vspace{-0.5cm}
\end{figure}

Next sections detail the main characteristics of the scenario with the aim of providing specific guidelines to drive the design of future MAC designs. As summarized in Figure \ref{fig:MACscenario}, we differentiate between {\em (i)} physical constraints, {\em (ii)} features of the on-chip network traffic, and {\em (iii)} requirements imposed by the multiprocessor architecture.

\subsection*{\centering \scshape \textbf{The Chip Scenario}} 
\label{sec:chipscenario}
We first focus our analysis on the physical constraints of the on-chip scenario.

\noindento
\emph{\textbf{Static and Controlled Landscape ---}} The propagation of electromagnetic waves takes place in a confined space. This physical landscape, including the network topology, the chip layout, and the characteristics of the employed materials, is fixed and known beforehand \cite{Matolak2012}. This represents one of the main uniquenesses of the \ac{WNoC} scenario, since nodes in other wireless networks generally move within a propagation environment that can also be dynamic. The intra-chip channel, in fact, becomes quasi-deterministic at the data-link layer and can be accurately characterized by exploiting \emph{a priori} knowledge of the physical landscape.

These considerations have profound implications on the MAC layer. Designers have a unique control over the transmission range, enabling one-hop communication and virtually eliminating problems such as {\em hidden} and {\em exposed} terminals. Therefore, techniques employed to address these issues at the cost of complexity and performance, e.g. RTS/CTS, are not necessary. This also opens the door to the use adaptive methods that require consensus to work, simplifying distributed protocols. Such control over propagation may also enable the detection of collisions, a functionality normally restricted to wired networks, which could greatly enhance the performance of random access protocols.


\noindento
\emph{\textbf{High Density of Nodes ---}} \acp{CMP} are expected to reach stunning levels of integration, enabling the development of thousand-core processors~\cite{Borkar2007}. This results in a wireless network where tens to hundreds of wireless nodes communicate simultaneously, context that sets {\em scalability} as a primary design objective. Although high-density wireless networks are not new, e.g. \acp{WSN}~\cite{Akyildiz2002}, nodes in such networks typically decrease their transmission range to reduce contention and resort to multiple hops to reach the intended destinations. However, this approach is not advisable here due to the stringent latency requirements of the application.

Scalability requirements also limit the practicality of channelization mechanisms. Creating tens to hundreds of orthogonal channels with the physical constraints of manycore chips is unfeasible due to the increase of complexity of key hardware, e.g., passband filters in frequency multiplexing or synchronization components in code multiplexing. Some works have tried to alleviate this issue by combining different multiplexing mechanisms, e.g. frequency and time in \cite{DiTomaso2015}. However, problems associated with the intrinsically rigid nature of the approach still arise whenever traffic conditions vary. Consider, as a specific yet relevant example, the poor performance of  multiplexing with a fair distribution of bandwidth in the presence of hotspot traffic. 

Scalability is also an issue in scheduling or random access protocols. In the former case, scheduling nodes need to manage requests from an increasing number of nodes and can easily become a performance bottleneck. In the latter case, designers can expect an increase of the collision probability as more nodes contend for the channel. Also, acknowledging becomes challenging in \acp{WNoC} devised to transport multicast traffic due to the expected burst of responses known as the {\em ACK implosion.}

\noindento
\emph{\textbf{Energy Awareness ---}} Typically nodes in wireless networks are mobile and hence have a limited battery. As a result, a large amount of research has been devoted to the development of protocols that are energy-efficient or even energy-constrained in extreme cases such as \acp{WSN}. In chip environments, energy availability is guaranteed, yet energy cannot be considered unlimited since heat dissipation is expensive. Actually, power has become a driver of multiprocessor design, suggesting the use of \ac{DVFS} techniques or power-gating techniques to increase the overall efficiency. \acp{WNoC} are also part of this design trend as they have been conceived to reduce the energy of long-range and collective transmissions, among other reasons.

This basically implies that MAC protocols for \ac{WNoC} need to be {\em energy-aware} in order to avoid diluting the energy benefits of wireless on-chip links. However, energy awareness should not come at the cost of an excessive drop of the network performance. In fact, latency constraints discourage the use of techniques with low duty cycles as proposed in numerous MAC protocols for \acp{WSN} \cite{Akyildiz2002}. Instead, the protocol should focus on reducing the protocol overhead and minimizing the penalty of collisions. Additionally, energy could be opportunistically saved by tuning --not turning off-- certain parts of the transceiver if a node is not expected to transmit in a while, e.g. during a backoff. Consistently with the \ac{DVFS} paradigm, protocols should also to adapt to changes in the frequency of cores as these may introduce variations not only in the traffic characteristics, but also on the performance of the wireless link. Finally, the correct operation of the protocol must not be disrupted if a given core, including its antenna and transceiver, is powered off. A simple example would be token passing, where the token could be lost when reaching the core wireless unit that has been shut down.


\noindento
\emph{\textbf{Underlying Infrastructure ---}} The \ac{WNoC} paradigm does not aim to replace wired on-chip networks, but rather to complement them. As a consequence, it is reasonable to assume the existence of underlying wired networks that provide a synchronized clock and can efficiently transport unicast flows. Such a wired backbone is unique to this scenario and it can be used to assist the MAC protocol by either transporting control information for scheduling or handshaking purposes, or absorbing traffic originally intended for the wireless network in congestion situations. For instance, our previous work in \cite{AbadalTPDS} discusses the use of a lightweight wired plane in token passing schemes to increase their scalability. Synchronization at the processor clock granularity can also be appealing when implementing slotted protocols. 

\subsection*{\centering \scshape \textbf{Workload Characteristics}} 
\label{sec:traffic}
Now, let us focus on the characteristics of the on-chip traffic to determine the flexibility required in multiprocessor networks.

\begin{figure*}[!t]
\centering
\subfigure[\label{mcastScalab}Percentage of long-range traffic (3 hops or more, top chart) and multicast traffic (bottom chart) for the SPLASH2 suite over MESI coherence \cite{AbadalCAEE}.]{\subfigtopskip = 0pt \subfigbottomskip = -15pt \includegraphics[width=0.32\textwidth]{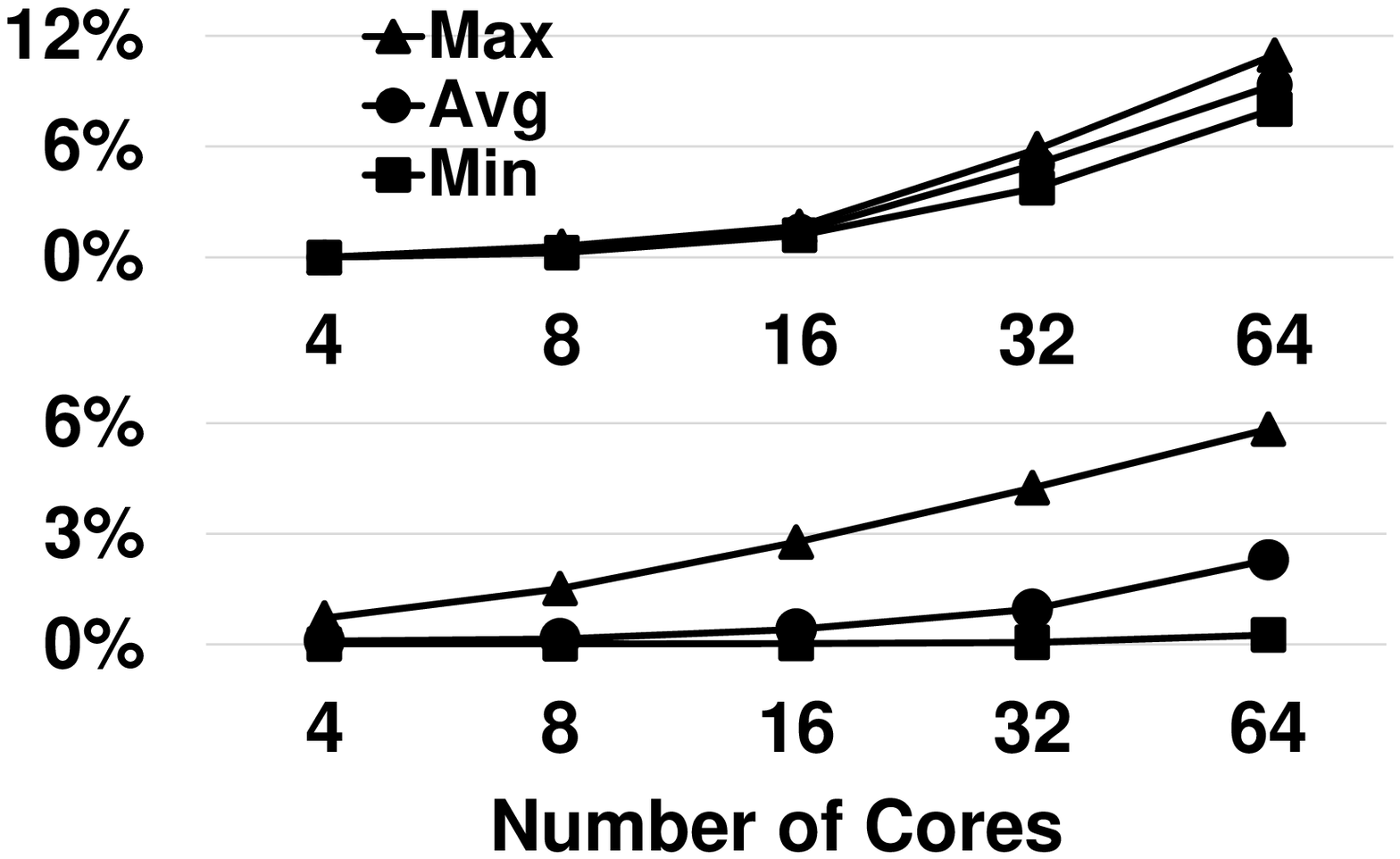}}
\subfigure[\label{phases}Phase behavior exhibited by the traffic generated by a 64-threaded instance of the {\em fluidanimate} application over MESI coherence \cite{AbadalCAEE}.]{\subfigtopskip = 0pt \subfigbottomskip = -15pt \includegraphics[width=0.32\textwidth]{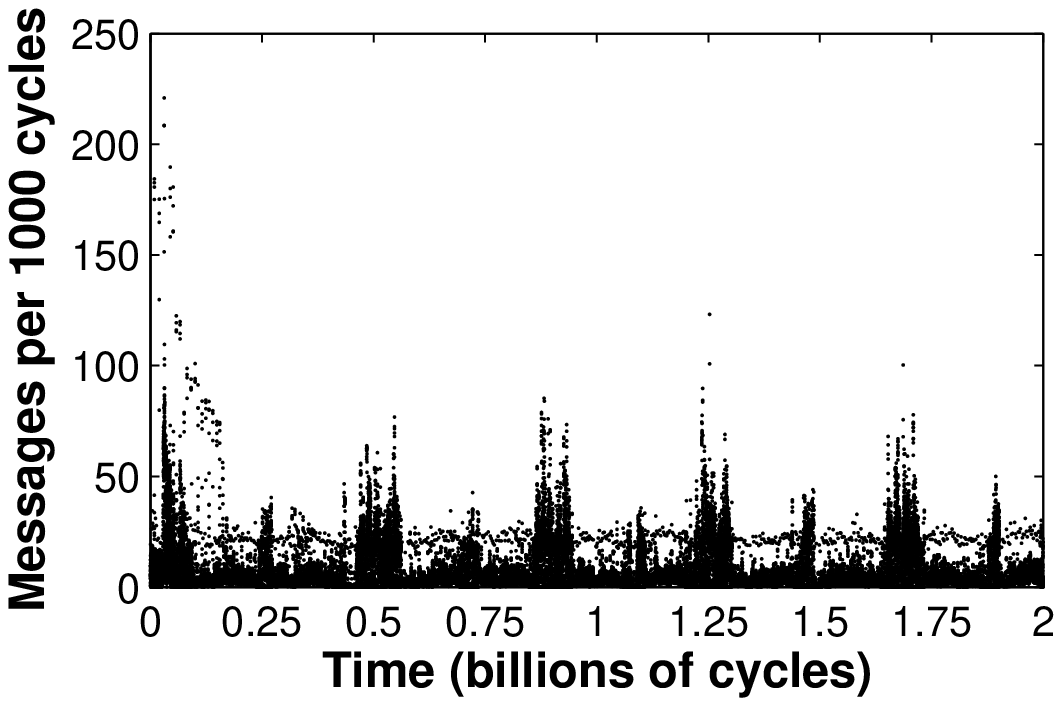}}
\subfigure[\label{Hsigma}Spatiotemporal characteristics of the applications analyzed in \cite{Soteriou2006}. Each mark represents a specific application.]{\subfigtopskip = 0pt \subfigbottomskip = -15pt \includegraphics[width=0.32\textwidth]{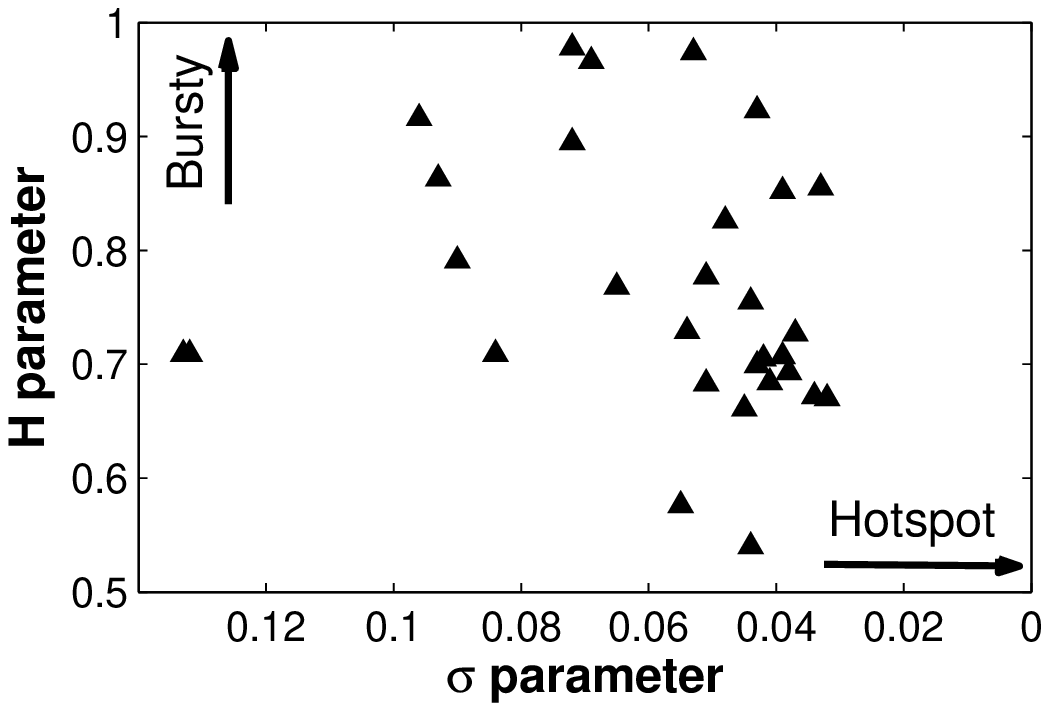}}
\vspace{-0.2cm}
\caption{Workload characterization of different multiprocessor architectures and applications exhibiting (a) increasing heterogeneity, (b) intra-application variability, and (c) inter-application variability with bursty and hotspot traffic.}
\vspace{-0.4cm}
\end{figure*}

\noindento 
\emph{\textbf{Heterogeneity and Variability ---}} Although architectures are generally designed trying to avoid expensive communication transactions, manycore processors face the challenge of having to support heterogeneous traffic profiles. Local and unicast communications dominate, but different works have demonstrated that the presence of global and multicast flows can become significant in manycore processors \cite{AbadalMICRO, Soteriou2006}. Figure \ref{mcastScalab} illustrates this by plotting the percentage of long-range and multicast traffic as a function of the number of cores with data extracted from our previous work in \cite{AbadalCAEE}. These are the types of traffic targeted by the \ac{WNoC} paradigm. 

The aforementioned works also discuss the variability of on-chip traffic. The existence of a wide range of programming models or application domains causes large changes in terms of communication demands from one application to another. Within each particular application, phase behavior also leads to wild variations on the traffic characteristics over time \cite{Sherwood2003}. Such a behavior is exemplified in Figure \ref{phases}, which clearly shows how the application {\em fluidanimate} alternates between communication-intensive and computation-intensive phases.  

The variability of traffic suggests that the MAC protocol should be reconfigurable to adapt to large-scale changes with a reasonable cost. This encourages the use of schemes that can be reconfigured periodically \cite{DiTomaso2015} or co-design techniques capable of reducing the uncertainty of intra- and inter-application variations. For instance, application phase prediction \cite{Sherwood2003} could be employed to foresee future traffic requirements and reconfigure the MAC protocol accordingly.

\noindento 
\emph{\textbf{Hotspot and Bursty Traffic ---}} As the predominant programming model in \acp{CMP}, shared memory has been assumed in the vast majority of \ac{NoC} works. Thus, coherence traffic has been characterized for a set of popular architectures and benchmarks with the aim to drive NoC evaluations and subsequent optimizations. The pioneering work by Soteriou {\em et al} revealed that, in most applications, a large fraction of the traffic is not only generated by a rather small subset of cores, but also injected in bursts \cite{Soteriou2006}. A similar workload characterization performed in our previous work \cite{AbadalCAEE} confirmed that multicast flows, as a subset of all on-chip traffic, also follow strong hotspot and bursty distributions. 


Figure \ref{Hsigma} represents the burstiness and hotspot levels obtained in \cite{Soteriou2006} for a wide set of applications. On the one hand, burstiness is represented via the Hurst exponent $H \in [0.5,1]$, where $H = 0.5$ corresponds to memoryless traffic and large values indicate the presence of self-similarity. On the other hand, the spatial concentration is characterized by means of a Gaussian standard deviation $\sigma \in [0,\infty)$, where small and large values represent hotspot and rather uniformly distributed traffic, respectively.

These spatiotemporal characteristics are generally detrimental to performance and call for flexible solutions that can provide fast and fine-grained adaptivity. Ideally, protocols would devote, if required, all resources to alleviate hotspots or to absorb traffic injection bursts. Multiplexing schemes lack such capability and thus are highly suboptimal within this context. Random access and scheduling protocols are more malleable and can manage hotspot traffic fairly well, but still suffer of high collision rates in the presence of bursty communication. In the latter case, the performance drop could be alleviated by anticipating upcoming bursts and adapting the protocol to them. This could be achieved by identifying recurrent correlation patterns \cite{AbadalCAEE} or, as outlined next, exploiting the monolithic nature of the system. 


\noindento 
\emph{\textbf{Monolithic System ---}} A multicore processor is basically a monolithic system from the designers' point of view and often a proprietary solution. The team responsible of designing the system therefore has --a rough-- control over the entire architecture, from the physical implementation up to the compiler that outputs the code. This represents a big departure from traditional wireless systems where the nodes, the network stack, and the applications are designed and developed by different teams and often rely on open standards. 

{\em The monolithic nature of the system basically implies that protocols can be streamlined by entering into the design loop of the whole architecture, enabling an unprecedented level of optimization of MAC protocols.} For instance, the compiler determines --to a large extent-- the distribution of the operations that generate the on-chip traffic within a given application, somehow fixing what and when the nodes will transmit. This knowledge could be leveraged to anticipate potentially harming situations or even to avoid them using a set of new compiling rules. A similar approach can be thought from the privileged perspective of the programmer. Consider a message passing system, where the programmer explicitly defines communication using a library of primitives. The MAC layer can be highly optimized by finely tuning the protocol to each primitive, especially in those related to collective communication. Similarly, the programmer could be provided with a set of special instructions that can explicitly define the behavior of the protocol for a given program section that may exhibit hotspot or bursty behavior. 


\subsection*{\centering \scshape \textbf{Architectural Requirements}} 
\label{sec:archrequirements}
The multiprocessor context imposes a set of performance objectives that impact on the design of the \ac{MAC} protocol. We subsequently outline a representative selection of these objectives.

\noindento
\emph{\textbf{Latency Sensitivity ---}} The latency introduced by on-chip communication essentially delays the progress of the computation that lies in the critical path of the processor. Latency becomes, in fact, more limiting than throughput in \acp{NoC} for cache-coherent manycore processors \cite{AbadalTPDS}. This highlights the appeal of MAC protocols that prioritize latency over throughput while taking into account the energy considerations discussed above.

In wireless networks, latency has traditionally been reduced by increasing the data rate, coordinating the multiple hops towards the receiver, or optimizing the periods of contention. The application of the first design rule to the multiprocessor context discourages the use of techniques that avoid contention by performing a fixed division of bandwidth, and favors random access and scheduling protocols instead. The second rule does not apply here as one-hop wireless transmissions are expected. With respect to the third rule, it is worth noting that the multiprocessor scenario offers unique optimization opportunities arising from the monolithic nature of the system as discussed previously. This information can be used to minimize the collision probability by adopting optimal persistence, backoff, or reservation policies; or by prioritizing latency-critical messages as per architecture, compiler, or programmer recommendations.

\noindento
\emph{\textbf{Stability and Fairness ---}} A link layer providing an homogeneous and bounded latency is very appealing feature from an architectural perspective, as it potentially reduces the complexity of the upper layers by rendering design decisions easier to reason and verify. This first implies that the MAC strategy must be stable, meaning that its performance should not decrease sharply beyond the saturation load to avoid latency peaks. Random access and scheduling protocols need to be carefully reviewed due to this, whereas channelization techniques are generally stable. In either case, protocols can always resort to the underlying wired \ac{NoC} to maintain the load below the saturation point. 
 
In practice, latency guarantees are generally addressed by means of protocols with \ac{QoS} \cite{Suriyachai2012}, which can apply priority policies to minimize the average or maximum latency of certain transactions. This will affect fairness, another important aspect that concerns latency and that has to be reviewed considering the criticality of the different flows. In the on-chip scenario, knowledge on the application can drive \ac{QoS} techniques to reduce the latency of certain critical flows. At the MAC level, techniques such as sharing the backoff counter among all nodes can also help increase fairness.

\noindento 
\emph{\textbf{Reliability ---}} Multicore processors generally require reliable communications, although this depends on the specific architecture. On the one hand, most processors implement error control and correction schemes to provide high-level reliability, as seemingly minor errors may corrupt an entire computation. Within this context, a reliable MAC solution is desirable to minimize the performance reduction caused by link-level errors. On the other hand, recent times have seen the emergence of approximate computing, which relaxes the need for fully precise computations to provide much higher energy efficiency. 


Unlike in traditional wireless networks, \ac{WNoC} proposals generally consider a \ac{BER} commensurate with that of \ac{RC} wires, i.e. $\sim 10^{-15}$. As a result, the MAC layer can safely assume that errors in the reception of a message will be caused by collisions in virtually all cases. Assuming a contention-based protocol, the objective will be to adjust the effort spent in resolving collisions according to the specific combination of latency and reliability requirements of the architecture. 

An important aspect to consider here is whether the static and controlled environment allows for the proactive detection of collisions. This would bring the contention-based MAC solution closer to that of wired networks, e.g. Ethernet, opening the door to strategies such as negative acknowledging. The implications are twofold. First, the latency of the protocol is highly reduced as there is no need for burdensome timeout approaches. Second, reliable broadcast support can be achieved with a reasonable cost. Note that, in wireless networks, broadcasts are generally {\em best effort} due to the severe congestion that would be caused by the subsequent acknowledgments. A simple solution here would consist in the transmission of a {\em tone} signal upon detecting a collision. Tones would be detected by idle nodes, including the colliding transmitters, and interpreted as a negative acknowledgment. If this approach is not technically viable, one can also relay any acknowledgment to the wired network, for which many-to-one traffic optimizations have been developed.

\section*{\Large \textbf{MAC Design Implications and Key Challenges}} 
Table \ref{tab:challenges} summarizes the differences between traditional wireless networks and on-chip networks, to then outline the potential implications of such differences on the design of MAC protocols for WNoC. The rest of the section highlights some of these design implications and discusses the main challenges associated to them. 

\begin{table*}
  \centering
	  \caption{Differences between traditional wireless networks and the Wireless Network-on-Chip scenario in terms of MAC protocol design.}
	\footnotesize
	\setlength\tabcolsep{3pt}
	\begin{tabular}{| >{\centering\arraybackslash}m{0.7in} || >{\centering\arraybackslash}m{1.8in} | >{\centering\arraybackslash}m{1.8in} || >{\centering\arraybackslash}m{1.8in} | }
	  \hline 
   & \textbf{Conventional Wireless Networks}  & \textbf{Wireless Network-on-Chip Scenario}   & \textbf{MAC Design Implications in Wireless Network-on-Chip} \\ \hline  \hline
	\multicolumn{4}{|c|}{\multirow{2}{*}{{\bf Physical Constraints \cite{Borkar2007, AbadalMICRO, Matolak2012, Niu2015, Akyildiz2002}}}} \\ 
	\multicolumn{4}{|c|}{} \\ \hline 
  Frequency              & Up to 60 GHz                      & 60 GHz and beyond & \multirow{2}{1.8in}{\centering Potentially high impact of propagation delay}  \\ \cline{1-3}
	Distance							 & From meters to kilometers			& A few centimeters   &     \\ \hline  
  Landscape              & Dynamic environment                 & Confined and static   &  Known range, consensus is easy to achieve   \\   \hline  
  \shortstack{Node\\Density}           & Moderate							 & High 								&	 Emphasis on scalability, channelization is discouraged  \\  \hline  
  \shortstack{Energy\\Policy}          & Case-dependent   & Limited by dissipation: energy-aware  & Reduce overhead or penalty of collisions    \\   \hline  
	\shortstack{Auxiliary\\Network}			 & Unreliable and expensive							 & Wired, reliable, and relatively cheap	&  Can be used for control, or prevent saturation \\ \hline \hline
  
	\multicolumn{4}{|c|}{\multirow{2}{*}{{\bf Workload Characteristics \cite{AbadalCAEE, Soteriou2006, Sherwood2003}}}} \\ 
	\multicolumn{4}{|c|}{} \\ \hline 
	\shortstack{System\\Design}         & Multi-vendor standardized system 	& Monolithic system		& Knowledge on traffic, co-design with upper layers and/or prediction  \\  \hline  
	\shortstack{Packet\\Length} 				 & Variable							              & Fixed, typically short   &  Nodes know when transmissions finish, can help fairness   \\   \hline  
	Broadcast              & Few sources, generally unreliable  & Any node can broadcast, needs to be reliable   &  Unified broadcast domain is preferable, scalable ACK  \\ \hline  
	Variability 				 & Due to traffic aggregation					& Phase behavior; difference between applications  & Reconfigurability is desirable \\ \hline 
	\shortstack{Spatiotemporal\\Characteristics} & Depends on context, spatiotemporal correlations may exist & Often bursty and hotspot & Adaptivity to fast changes is desirable \\ \hline \hline

		\multicolumn{4}{|c|}{\multirow{2}{*}{{\bf Architectural Requirements \cite{AbadalTPDS, Suriyachai2012, Akyildiz2002}}}} \\ 
		\multicolumn{4}{|c|}{} \\ \hline 
  Latency                & Variable, generally not a strong requirement            & May be critical   & \ac{QoS}-aware design principles are desired  \\ \hline    
	Throughput						 & Generally important								 & Important but not critical		&  Give priority to latency over throughput \\ \hline  
	Reliability            & Errors can be tolerated   					 & Errors are mostly not tolerated  & Strong emphasis on reliability     \\  \hline 
   
  \end{tabular}

  \label{tab:challenges}
\end{table*}

The on-chip scenario is driven by latency and reliability, yet with a strong emphasis on energy efficiency. This combination of requirements may have been, together with simplicity, the main cause of the relatively widespread proposal of multiplexing \cite{Deb2013}. However, it seems apparent that scalability constraints or flexibility requirements will lead to scheduling or random access solutions instead \cite{AbadalTPDS}. We believe that the sweet spot is somewhere in between these two extremes, in schemes that combine the advantages of time multiplexing, i.e. high throughput and fairness; and of random access, i.e. low latency and flexibility. The challenge here is to develop solutions capable of balancing performance and flexibility effectively. A first attempt is made in \cite{Mansoor2015} by implementing both token passing and \ac{CSMA} and designing a controller that dynamically switches between both strategies depending on the load. A step further towards this direction would be the use of protocols that naturally and gradually adapt their characteristics to the load without the need of an external controller. The literature contains a wealth of proposals for conventional networks that achieve such a functionality by employing unconventional persistence mechanisms, random access with reservation, or probabilistic time division. Future works could evaluate the applicability of these schemes in the WNoC scenario and explore their performance--flexibility tradeoffs. 

The waste of time and energy caused by collisions is most likely the reason why random access techniques have been scarcely proposed thus far. However, the chip scenario offers a unique opportunity for the practical implementation of collision detection, a major breakthrough in any wireless network, as well as the enabler of strategies to reduce the penalty of collisions. Obviously, the main technical challenges here are to demonstrate the feasibility of collision detection in this scenario and under which conditions. Also, the transmitter will probably not be able to perform the detection as received signals will be masked by much stronger transmitted signals and, therefore, protocols may need to adapt to this fact. In \cite{Mestres2016}, we explore different ways to address these challenges. We first propose to take advantage of the time-invariant and known propagation medium to effectively detect collisions. For instance, collisions could be detected by the receivers by comparing the expected received power (which can be calculated \emph{a priori} using the source address of a packet) and the actual received power. Then, we propose a streamlined collision notification scheme where receivers broadcast a jamming signal as soon as they detect a collision. On the other hand, transmitters initially send a preamble instead of a full packet, check for the jamming signal, and then transmit the rest of the message if the jamming signal is absent. Otherwise, they back off \cite{Mestres2016}. 

Besides collision detection, the main particularity of the WNoC scenario is its monolithic nature. Knowledge on the architecture and the application being run enables the use of prediction to improve the performance and efficiency of the MAC protocol. Predictive schemes are in fact pervasive in current architectures as they are employed to optimize the processor pipeline, the coherence protocol, or the wired \ac{NoC} at runtime. Before applying such techniques in WNoC, however, it is necessary to {\em (i)} identify recurrent correlation patterns to exploit, {\em (ii)} design fast and compact predictors capable of capturing them, and {\em (iii)} integrate the predictor within the MAC protocol in an optimal way. As a first approximation, we recently implemented a last value predictor that anticipates the source of the next wireless transmissions \cite{AbadalCAEE}. Our prediction obtained an accuracy of up to 80\% in parallel computing benchmarks and can be used, for instance, to eliminate a large portion of performance-degrading collisions in random access protocols. However, further research is required to fully demonstrate the potential of an approach where compiler-assisted techniques or phase prediction and tracking can be leveraged to improve performance even further.




\section*{\Large \textbf{Conclusion}} 
\label{sec:conclusions}
With the potential to offer low-power and low-latency global and broadcast communication, wireless on-chip communication holds great promise for the implementation of \acp{NoC} for manycore chips. However, it is necessary to develop scalable, fast, and efficient \ac{MAC} mechanisms to exploit such potential. This position paper has provided a rigorous context analysis with the aim to clarify the singularities of this new environment for \ac{MAC} protocol research, which features both very stringent requirements and unique optimization opportunities. On the one hand, the analysis highlights latency, reliability, and variability of traffic as the most challenging aspects of the scenario; on the other hand, the analysis points to the static, controlled, and monolithic nature of a multiprocessor as the most salient characteristics potentially leading to unprecedented performance via cross-layer design far beyond traditional limits. Indeed, prediction and other MAC-architecture co-design techniques are unique to the multiprocessor scenario and open a large and exciting design space to be explored in future works.

\section*{\Large \textbf{Acknowledgment}} 
This work has been supported in part by NSF under grant CCF 16-29431, the European Commission under grant H2020-FETOPEN-736876 (VISORSURF), the Spanish MINECO under contract TEC2017-90034-C2-1-R (ALLIANCE) that receives funding from FEDER, and by the Catalan Institution for Research and Advanced Studies (ICREA).

\end{document}